\documentclass[lineno]{jfm}

\usepackage{graphicx}
\usepackage{newtxtext}
\usepackage{newtxmath}
\usepackage{natbib}
\usepackage{mathtools}
\usepackage{hyperref}
\usepackage{tikz}
\usepackage{cleveref}
\hypersetup{
    colorlinks = true,
    urlcolor   = blue,
    citecolor  = blue,
}

\newcommand{\wT}{\mbox{$\smash{\mean{wT}}$}}

\newcommand{\dT}{ \smash{{\horav{\delta T}}}  }

\newcommand{\volav}[1]{\left\langle #1 \right\rangle}

\newcommand{\horav}[1]{\langle #1 \rangle_h}

\newcommand{\mean}[1]{{\volav{#1}}}

\newcommand{\subeqref}[2]{\hyperref[#1]{(\ref*{#1}#2)}}

\definecolor{matlabblue}{RGB}{0,113,188}
\definecolor{matlabred}{RGB}{216,82,24}
\definecolor{mygrey}{rgb}{0.7,0.7,0.7}
\definecolor{matlabgreen}{rgb}{0,0.498,0} 
\definecolor{black}{rgb}{0,0,0}
\definecolor{colorbar1}{rgb}{1.000000,0.909091,0.000000}
\definecolor{colorbar2}{rgb}{1.000000,0.818182,0.000000}
\definecolor{colorbar3}{rgb}{1.000000,0.727273,0.000000}
\definecolor{colorbar4}{rgb}{1.000000,0.636364,0.000000}
\definecolor{colorbar5}{rgb}{1.000000,0.545455,0.000000}
\definecolor{colorbar6}{rgb}{1.000000,0.454545,0.000000}
\definecolor{colorbar7}{rgb}{1.000000,0.363636,0.000000}
\definecolor{colorbar8}{rgb}{1.000000,0.272727,0.000000}
\definecolor{colorbar9}{rgb}{1.000000,0.181818,0.000000}
\definecolor{colorbar10}{rgb}{1.000000,0.090909,0.000000}
\definecolor{colorbar11}{rgb}{1.000000,0.000000,0.000000}
\definecolor{colorbar12}{rgb}{0.909091,0.000000,0.000000}
\definecolor{colorbar13}{rgb}{0.818182,0.000000,0.000000}
\definecolor{colorbar14}{rgb}{0.727273,0.000000,0.000000}
\definecolor{colorbar15}{rgb}{0.636364,0.000000,0.000000}
\definecolor{colorbar16}{rgb}{0.545455,0.000000,0.000000}
\definecolor{colorbar17}{rgb}{0.454545,0.000000,0.000000}
\definecolor{colorbar18}{rgb}{0.363636,0.000000,0.000000}
\definecolor{colorbar19}{rgb}{0.272727,0.000000,0.000000}
\definecolor{colorbar20}{rgb}{0.181818,0.000000,0.000000}
\definecolor{colorbar21}{rgb}{0.090909,0.000000,0.000000}

\definecolor{matlabyellow}{rgb}{0.93,0.69,0.13} 

\newcommand\solidrule[1][10pt]{\rule[0.5ex]{#1}{1.5pt}}
\newcommand\dashedrule{\mbox{\solidrule[2pt]\hspace{2pt}\solidrule[2pt]\hspace{2pt}\solidrule[2pt]}}

\newcommand{\mycross}[1]{%
	\protect\begin{tikzpicture}%
	\protect\draw[thick,color=#1] (0,0) -- (1ex,1ex);
	\protect\draw[thick,color=#1] (0,1ex) -- (1ex,0);
	\protect\end{tikzpicture}%
}

\renewcommand{\vec}[1]{\mathbf{#1}}

\renewcommand{\theta}{\vartheta}
\renewcommand{\phi}{\varphi}

\newcommand{\RomanNumeralCaps}[1]

\title{
Internal heating profiles for which downward conduction is impossible
}

\author{Ali Arslan\aff{1}
  \corresp{\email{ali.arslan@erdw.ethz.ch}},
 Giovanni Fantuzzi\aff{2}, John Craske\aff{3}
 \and Andrew Wynn\aff{4}}

\affiliation{\aff{1}
Institute of Geophysics, ETH Z\"{u}rich, Z\"{u}rich, CH-8092, Switzerland
\aff{2}Department of Mathematics, FAU Erlangen-N\"{u}rnberg, Erlangen, 91054, Germany
\aff{3} Department of Civil and Environmental Engineering, Imperial College London, London, SW7 2AZ, UK
\aff{4} Department of Aeronautics, Imperial College London, London, SW7 2AZ, UK
}

\begin{document}
\nolinenumbers
\maketitle

\nolinenumbers
\begin{abstract}

\noindent

We consider an internally heated fluid between parallel plates with fixed thermal fluxes. For a large class of heat sources that vary in the direction of gravity, we prove that $\smash{\dT} \geq \sigma R^{-1/3} - \mu$, where $\smash{\dT}$ is the average temperature difference between the bottom and top plates, $R$ is a `flux' Rayleigh number and the constants $\sigma,\mu >0$ depend on the geometric properties of the internal heating. This result implies that mean downward conduction (for which $\smash{\dT}< 0$) is impossible for a range of Rayleigh numbers smaller than a critical value $R_0$. The bound demonstrates that $R_0$ depends on the heating distribution and can be made arbitrarily large by concentrating the heating near the bottom plate. However, for any given fixed heating profile of the class we consider, the corresponding value of $R_0$ is always finite. This points to a fundamental difference between internally heated convection and its limiting case of Rayleigh-B\'enard convection with fixed flux boundary conditions, for which $\dT$ is known to be positive for all $R$.

\end{abstract}


\section{Introduction}
\label{sec:intro}

Convection driven by spatially varying heating is attracting attention as a generalisation of internally heated convection (IHC) due to its relevance in studying different regimes of heat transport by turbulence. Recent work demonstrates that varying the heating location can enhance heat transport in bounded domains \citep{kazemi2022,bouillaut2022, Lepot2018} and that bounds on the heat transport depend on the supply of potential energy due to variable heating \citep{Song2022}. 
Such studies of non-uniformly heated convection have implications for geophysical fluid dynamics, from mixing in lakes, where solar radiation acts as a spatially varying heat source, to convection in the Earth's mantle and liquid outer core driven by energy released irregularly by radioactive isotopes and secular cooling \citep{schubert2015treatise}. Furthermore, non-uniform heating is theoretically significant because it induces a much larger class of flows than those by boundary-forced examples such as Rayleigh-B\'enard convection (RBC).

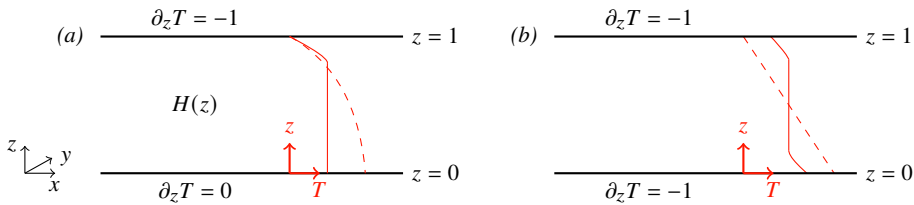
\begin{figure}
    \centering
    \hspace{-1cm}
    \begin{tikzpicture}[every node/.style={scale=0.95}]
    \draw[black,thick] (-6,0) -- (-2,0) node [anchor=west] {$z=0$};
    \draw[black,thick] (-6,1.8) -- (-2,1.8) node [anchor = west] {$z=1$};
    \draw [colorbar10] plot [smooth, tension = 1] coordinates {(2.86,1.8) (3.02,1.62) (3.1,1.5)};
    \draw [colorbar10] plot [smooth, tension = 1] coordinates {(3.1,0.3) (3.16,0.18) (3.34,0)};
    \draw [dashed ,colorbar10] (2.5,1.8) -- (3.7,0);
    \draw [colorbar10] (3.1,1.5) -- (3.1,0.3);
    \node at (1.25,-0.25) {$ \partial_z T = -1 $};
    \node at (1.25,2.025) {$ \partial_z T = -1 $};
    \node at (-4.75,0.9) { $H(z)$ };
    \draw[->] (-7,0) -- (-7,0.36) node [anchor=east]{$z$};
    \draw[->] (-7,0) -- (-6.6,0) node [anchor=north]{$x$};
    \draw[->] (-7,0) -- (-6.64,0.2) node [anchor=west]{$y$};
    \draw[black,thick] (0,0) -- (4,0) node [anchor=west] {$z=0$};
    \draw[black,thick] (0,1.8) -- (4,1.8) node [anchor = west] {$z=1$};
    \node at (-4.75,-0.25) {$ \partial_z T = 0 $};
    \node at (-4.75,2.025) {$  \partial_z T = - 1 $};
    \draw [dashed,colorbar10] plot [smooth, tension = 1] coordinates {(-3.5,1.8) (-2.8,1.1) (-2.5,0)};
    \draw [colorbar10] plot [smooth,tension=1] coordinates {(-3.5,1.8) (-3.12,1.58) (-3,1.45)}; 
    \draw [colorbar10] (-3,1.45) -- (-3,0);
    \draw[->,colorbar10,thick] (2.5,0) -- (2.5,0.4) node [anchor=south]{$z$};
    \draw[->,colorbar10,thick] (2.5,0) -- (2.9,0) node [anchor=north]{$T$};
   \draw[->,colorbar10,thick] (-3.5,0) -- (-3.5,0.4) node [anchor=south]{$z$};
    \draw[->,colorbar10,thick] (-3.5,0) -- (-3.1,0) node [anchor=north]{$T$};
    \node at (-6.4,1.8) {\textit{(a)}};
    \node at (-0.4,1.8) {\textit{(b)}};
    \end{tikzpicture}
    \caption{\textit{(a)} Internally heated and \textit{(b)}  Rayleigh-B\'enard convection with fixed flux boundaries, where $H(z)$ is a positive non-uniform heating profile. In both panels, ({\color{colorbar10}\dashedrule}) denotes the conductive temperature profiles and ({\color{colorbar10}\solidrule}) the mean temperature profiles in the turbulent regime. In panel \textit{(a)}, the temperature profiles are for $H(z)=1$. }
    \label{fig:config}
\end{figure}

This work considers non-uniform IHC between parallel plates with fixed thermal fluxes. Specifically, we assume that the lower plate is a thermal insulator and that the upper plate is a poor conductor in comparison to the fluid's ability to transport heat \citep[see, for example,][]{Goluskin2015a}. The heat flux through the upper plate is therefore assumed fixed  and, to ensure that a statistically stationary state is realisable, is specified to match the total heat input due to internal heating (\cref{fig:config}\textit{(a)}). The Rayleigh number $R$ for this system, defined precisely in \S\ref{sec:setup}, quantifies the destabilising effect of internal heating relative to the stabilising effect of diffusion. An emergent physical property of the flow on which we will focus is the mean temperature difference, $\dT$, between the lower and upper plates (we use $\volav{\cdot}$ to represent an infinite time and volume average, with subscript $h$ denoting a horizontal average).

This paper communicates bounds on $\dT$ that depend on the Rayleigh number $R$ of the flow and the spatial distribution of the heat sources. More specifically, we prove that for any solution of the governing Boussinesq equations,
\begin{equation}
    \label{eq:main_res}
    \dT := \horav{T|_{z=0}} - \horav{T|_{z=1} } \geq \sigma R^{-\gamma}-\mu, 
\end{equation}
where $\sigma,\mu>0$ are constants depending only on the spatial distribution of the internal heating, and $\gamma$ is a scaling rate.  A corollary of \eqref{eq:main_res} is that 
downward conduction, here defined as the emergent mean temperature of the upper plate being above that of the lower plate, i.e., $ \horav{T|_{z=1}} > \horav{T|_{z=0}}$, is impossible for $R\leq R_0= (\sigma/\mu)^{1/\gamma}$, where $R_0$ is a critical Rayleigh number that depends on the spatial distribution of the internal heating.

To give context to this result, observe that a flow for which a rigorous characterisation of $\dT$ is known is fixed-flux RBC, shown schematically in \cref{fig:config}\textit{(b)}. In that case, symmetric heating and cooling rates are prescribed at the upper and lower boundaries (as opposed to internally), and the temperature drop satisfies the rigorous bound $\dT \geq cR^{-1/3}$ for a constant $c>0$ \citep{Otero2002}. The physical interpretation of this result is that the upper plate, through which heat leaves the domain, must, on average, remain colder than the lower one. Further, the difference in temperature between the plates cannot decrease at a rate faster than $R^{-1/3}$, where $R \rightarrow \infty$ corresponds to an increasingly turbulent and homogenised flow. Noting that $R$ is a flux-based Rayleigh number (rather than being defined in terms of an imposed temperature difference), the provable decay rate of the lower bound on $\dT$ agrees with the predictions of some phenomenological theories, namely the so-called ultimate regime \citep{kraichnan1962,spiegel1963generalization}. See \cite{  doering2020turning} and \cite{lohse2023} for recent reviews on the ultimate regime in RBC.

There is a striking and crucial distinction between what can currently be proven about the behaviour of the temperature drop $\langle \delta T \rangle_h$ in IHC in comparison with fixed-flux RBC. In particular, while for fixed-flux RBC downward conduction can be ruled out for all Rayleigh numbers (i.e., that $\dT >0$ must hold for all $R$), it is not yet known whether there are any heating profiles $H(z)$ that lead to the equivalent property in IHC. For example, in the case where internal heating is applied uniformly ($H \equiv 1$), it was shown by \cite{Arslan2021a} that $\dT \geq 1.6552R^{-1/3}-0.03868$. This only rules out downward conduction up to the critical Rayleigh number  $R_0 = 78\,389 = 54.437 R_L$, where $R_L=1440$ denotes the Rayleigh number at which the flow becomes linearly unstable \citep{Goluskin2015a}. 

\begin{figure}
    \centering
    \hspace{-1cm}
    \begin{tikzpicture}[every node/.style={scale=0.95}]
    \fill [colorbar14] (-7,1.8) rectangle (-3.5,1.2);
    \fill [colorbar14] (-3,0) rectangle (0.5,0.6);
    \shade [top color = colorbar6, bottom color = colorbar13] (1,0.9) rectangle (4.5,0.45);
    \draw [colorbar13] (1,0.45) -- (4.5,0.45);
    \shade [top color = colorbar13, bottom color = colorbar6] (1,0.45) rectangle (4.5,0);
    \draw [colorbar6] (1,0.9) -- (4.5,0.9);
    \shade [top color = colorbar6, bottom color = colorbar1] (1,1.8) rectangle (4.5,1.35);
    \draw [colorbar1] (1,1.35) -- (4.5,1.35);
    \shade [top color = colorbar1, bottom color = colorbar6] (1,1.35) rectangle (4.5,0.9);
    \draw[black,line width=0.35mm] (-7,0) -- (-3.5,0) ;
    \draw[black,line width=0.35mm] (-7,1.8) -- (-3.5,1.8) ;
    \draw[black,line width=0.35mm] (-3,0) -- (0.5,0) ;
    \draw[black,line width=0.35mm] (-3,1.8) -- (0.5,1.8) ;
    \draw[black,line width=0.35mm] (1,0) -- (4.5,0) node [anchor=west] {$z=0$};
    \draw[black,line width=0.35mm] (1,1.8) -- (4.5,1.8) node [anchor = west] {$z=1$};
    \draw plot [smooth, tension = 1] coordinates {(-4.8,1.8) (-4.2,1.5) (-4,1.2)};
    \draw (-4,1.2) -- (-4,0);
    \draw plot [smooth, tension = 1] coordinates {(0,0) (-0.12,0.3) (-0.5,0.6)};
    \draw (-0.5,0.6) -- (-0.8,1.8);
    \draw plot [smooth, tension = 1] coordinates {(3.2,1.8) (3.8,0.8) (4,0)};
    \node at (-7.2,1.8) {\textit{(a)}};
    \node at (-3.2,1.8) {\textit{(b)}};
    \node at (0.8,1.8) {\textit{(c)}};
    \end{tikzpicture}
    \caption{ Examples of non-uniform heating profiles in a non-dimensionalised domain with illustrative sketches of the conductive temperature profiles (\solidrule) for fixed flux boundary conditions. In \textit{(a)}, the heating is localised near the upper boundary and in \textit{(b)} near the lower boundary while being zero elsewhere, shown with red ({\color{colorbar14}\solidrule}) and white spaces respectively. For \textit{(c)} the heating is sinusoidal with a maximum at $z=0.25$ ({\color{colorbar13}\solidrule}) and a minimum of zero at $z=0.75$ ({\color{colorbar1}\solidrule}).   }
    \label{fig:config2}
\end{figure}
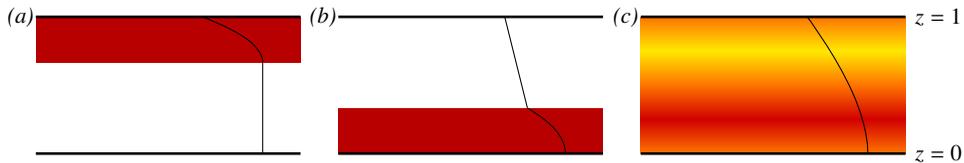

The bound \eqref{eq:main_res}, which is the main result of this paper, gives valuable additional information on the gap between the known behaviour of $\dT$ for IHC with an insulating bottom plate and fixed-flux RBC. Specifically, it allows us to understand how the gap depends on the spatial distribution $H(z)$ of the internally applied heating via the constants $\mu = \mu(H)$ and $\sigma = \sigma(H)$ in \eqref{eq:main_res}. The dependence of this gap on $H$ is interesting since it is not unreasonable to expect that downward conduction may be more likely to occur if heating is concentrated close to the upper plate, as in Figure \cref{fig:config2}~(a), rather than close to the lower plate, as in Figure \cref{fig:config2}~(b). The specific expressions for $\mu$ and $\sigma$ derived in this paper do not allow downward conduction to be ruled out (i.e. $R_0 <\infty$) for any {\em fixed} heating profile $H$. However, we show that it is possible to construct a sequence of heating profiles $H_r$, indexed by an integer $r$, in which heating is asymptotically concentrated towards the lower plate as $r \rightarrow \infty$, for which 
\[
R_0(H_r) = \left( \frac{\sigma(H_r)}{\mu(H_r)} \right)^{ \frac{1}{\gamma(H_r)}} \rightarrow \infty \quad \text{as} \quad   r\rightarrow \infty.
\]
In this sense, we can partially bridge the gap between what is known about $\dT$ for fixed-flux RBC and what appears, at least from the perspective of rigorous mathematical analysis, to be the much more intricate behaviour of internally heated convective flows.

In summary, this paper proves a Rayleigh-dependent lower bound on $\dT$ that depends on the geometric properties of the internally applied heating. As a corollary, for any heating profile, we can find an expression for the critical Rayleigh number $R_0$ below which downward conduction is impossible for any solution of the governing equations. To do this, we develop the mathematical approach of \cite{Arslan2021a} for flows driven by arbitrary (non-uniform) heating. The paper is structured as follows: \S\ref{sec:setup} describes the problem setup, in \S\ref{sec:Bounds} we prove the lower bound \eqref{eq:main_res}, \S\ref{sec:Imp_bound} highlights implications of the bound on the possibility of downward conduction in geophysical flows and in \S\ref{sec:conclusion} we conclude.

\section{Setup}
\label{sec:setup}

We consider a layer of fluid between two horizontal plates separated by a distance $d$ and periodic in the horizontal ($x$ and $y$) directions with periods $L_x d$ and $L_y d$. The fluid has kinematic viscosity $\nu$, thermal diffusivity $\kappa$, density $\rho$, specific heat capacity $c_p$ and thermal expansion coefficient $\alpha$. Gravity acts in the negative vertical direction, and the fluid is heated internally at a position-dependent non-negative volumetric rate $\Tilde{H}\geq 0$. We assume that $\Tilde{H}$ is a non-negative integrable function on the domain that depends only on the vertical coordinate and satisfies   $\langle \tilde{H}\rangle>0$ strictly, and define a dimensionless heating as $H(z):=\tilde{H}/\langle \tilde{H}\rangle$. Note that $\langle H\rangle=\|H\|_{1}=1$ because $H(z)\geq 0$. Here and throughout the paper, $\lVert f \rVert_p$ is the standard $L^p$ norm of a function $f:[0,1]\rightarrow\mathbb{R}$.

To non-dimensionalise the problem, we use $d$ as the characteristic length scale, $d^2/\kappa$ as the time scale and $d^2\langle \tilde{H}\rangle/\kappa \rho c_p$ as the temperature scale. The velocity of the fluid $\boldsymbol{u}(\boldsymbol{x},t)=(u(\boldsymbol{x},t),v(\boldsymbol{x},t),w(\boldsymbol{x},t))$ and temperature $T(\boldsymbol{x},t)$ in the non-dimensional domain $\Omega = [0,L_x]\times[0,L_y]\times[0,1]$ are then governed by the Boussinesq equations,
\begin{subequations}
\label{eq:gov_eqs}
\begin{align}   
    \label{eq:continuit}
    \bnabla \cdot \boldsymbol{u} &= 0,\\
    \label{eq:mom_eq}
    \partial_t \boldsymbol{u} + \boldsymbol{u} \cdot \bnabla \boldsymbol{u} + \bnabla p &= Pr\, \bnabla^2 \boldsymbol{u} + Pr\, R\, T \vec{e}_3 ,   \\
    \label{eq:nondim_energy_nonUniform}
    \partial_t T + \boldsymbol{u}\cdot \bnabla T &= \bnabla^2 T + H(z).
\end{align}
\end{subequations}
The non-dimensional numbers are the Prandtl and Rayleigh numbers, defined as 
\begin{equation}
    Pr = \frac{\nu}{\kappa},
    \qquad\text{and}\qquad 
    R = \frac{g \alpha \langle \tilde{H}\rangle d^5 }{\rho c_p \nu \kappa^2 }.
\end{equation}
The boundary conditions are no-slip for the velocity and fixed flux for the temperature:
\begin{subequations}
\label{bc:non_uni}
\begin{gather}
    \label{bc:u_non_u}
    \boldsymbol{u}|_{z=\{0,1\}} = 0, \\
    \label{bc:nu}
    \partial_z T|_{z=0} = 0, \quad \partial_z T|_{z=1}=-1.
\end{gather}
\end{subequations} 
 A diagram of the system is shown in \cref{fig:config}\textit{(a)}.

It proves useful to define a function $\eta \in L^2(0,1)$ such that $\eta(z)$ measures the total (dimensionless) heat  added to those parts of the domain below a height $z$:
\begin{equation}
    \label{eq:eta_general}
    \eta(z):= \int^{z}_0 H(z)\, \textrm{d}z.
\end{equation}
The value $\eta(1)=1$ corresponds to the total heat added to the domain and, therefore, to the negative heat flux applied at the top boundary in \eqref{bc:nu}.
We can obtain an identity for the mean vertical heat transport of the system by multiplying \eqref{eq:nondim_energy_nonUniform} by $z$ and taking an infinite-time and volume average. Use of \eqref{eq:eta_general} and a standard application of integration by parts with the boundary conditions \eqref{bc:u_non_u} \& \eqref{bc:nu} gives the identity
\begin{align}   
\label{eq:energy_bal_nu}
    \wT + \dT &= -\int^{1}_{0}(z-1)H\,\textrm{d}z=\int^{1}_{0}\eta(z)\,\textrm{d}z.
\end{align}
The left-hand side of \eqref{eq:energy_bal_nu} is the sum of the mean vertical convective heat flux, $\wT$, and the mean conductive heat flux, $\dT$, and balances the potential energy added to the system. To see why, note that $-zH$ is the pointwise supply of (dimensionless) potential energy (the negative sign is there because potential energy is created by positive heating from below or negative heating from above). The middle term in \eqref{eq:energy_bal_nu} therefore accounts for the potential energy created by internal heating and cooling from the top boundary (alternatively, the shifted coordinate $z\mapsto z-1$, centred on the top boundary, can be seen as a means of removing the cooling at the top boundary from the calculation).

To remove the inhomogeneous boundary conditions on $T$, it is convenient to rewrite the temperature field in terms of perturbations $\theta$ from the conductive profile $T_c$,
\begin{equation}
    \label{eq:T}
    T(\boldsymbol{x},t) = \theta(\boldsymbol{x},t)  + T_c(z).
\end{equation}
The steady conductive temperature profile can be found after taking $\boldsymbol{u}=0$ and $T(\boldsymbol{x},t)=T(\boldsymbol{x})$ in \eqref{eq:nondim_energy_nonUniform}. Given the insulating lower boundary condition, $T'_c(z) = - \eta(z)$, where primes denote derivatives with respect to $z$. Then, \eqref{eq:nondim_energy_nonUniform} in terms of $\theta$ becomes
\begin{gather}
    \label{eq:theta_eq}
     \partial_t \theta + \boldsymbol{u} \cdot \bnabla \theta = \bnabla^2 \theta + w\, \eta(z) ,\\
    \label{bc:theta_non_u}
    \partial_z \theta|_{z=\{0,1\}} = 0 \, .
\end{gather}

\section{Bounding heat transport}
\label{sec:Bounds}

To bound $\dT$ using methods that have been successfully applied to uniform IHC \citep{Arslan2021a}, we search for an upper bound on $\wT$ and use \eqref{eq:energy_bal_nu} to bound $\dT$. We will prove upper bounds on  $\mean{w\theta}$, which is equal to $\wT$ because the incompressibility of the velocity field and the boundary conditions \eqref{bc:u_non_u} imply that any function $f:[0,1]\rightarrow\mathbb{C}$ satisfies $\volav{w f} = \int^{1}_0 \horav{w} f(z) \textrm{d}z = 0$. The derivation in  \cref{sec:afm} is analogous to that in \cite{Arslan2021a}, so we only outline the main ideas in this paper.

\subsection{The auxiliary functional method}
\label{sec:afm}

Rigorous bounds on the mean quantities of turbulent flows can be found with the background method \citep{Doering1994, Doering1996,Constantin1995a}, which we formulate here in the more general framework of the auxiliary functional method \citep{Chernyshenko2014a,Fantuzzi2022}. We consider the quadratic auxiliary functional
\begin{equation}
    \label{eq:af}
    \mathcal{V}\{\boldsymbol{u},\theta\} =  \fint_\Omega  \frac{a}{2\,Pr\,R} |\boldsymbol{u}|^2 + \frac{b}{2}\left|\theta  - \frac{\phi(z)}{b}\right|^2 \textrm{d}\boldsymbol{x} ,
\end{equation}
where $a$ and $b$ are non-negative scalars and $\phi(z)$ a function, all to be optimised. As demonstrated by \cite{Chernyshenko2022}, the use of quadratic auxiliary functionals is equivalent to the background method, where $a$ and $b$ are referred to as balance parameters and $\phi(z)/b$ is the background temperature field 
 satisfying the boundary conditions on $\theta$ in \eqref{bc:theta_non_u}. With a slight abuse of terminology, we refer to $\phi(z)$ as the background field. 

Given that $\mathcal{V}\{\boldsymbol{u},\theta\}$ remains bounded in time along solutions of \eqref{eq:continuit}, \eqref{eq:mom_eq} \& \eqref{eq:theta_eq} for any given initial $\boldsymbol{u}$ and $\theta$, the infinite-time average of the time derivative of $\mathcal{V}$ is zero. Using this property, we can write
\begin{equation}
    \label{eq:AFM_statemen}
    \mean{w\theta} = U - \left( U - \volav{w\theta} - \limsup_{\tau\rightarrow\infty} \frac{1}{\tau}\int^{\tau}_0\frac{\textrm{d}}{\textrm{d}t} \mathcal{V}\{\boldsymbol{u},\theta\} \textrm{d}t  \right),
\end{equation}
and deduce that $\mean{w \theta} \leq U$, if after rearranging, the functional in the brackets in \eqref{eq:AFM_statemen} is pointwise in time non-negative. 
Following computations analogous to \cite{Arslan2021a}, the terms in the brackets of \eqref{eq:AFM_statemen} becomes
\begin{equation}
    \label{eq:s_f}
    \mathcal{S}\{\boldsymbol{u},\theta\}:=\fint_{\Omega} \frac{a}{R} |\bnabla \boldsymbol{u}|^2 + \beta |\bnabla \theta|^2 - (a+1+b\eta(z) - \phi'(z))w\theta - \phi'(z)\partial_z\theta + U \, \textrm{d}\boldsymbol{x}.
\end{equation}
The positivity of \eqref{eq:s_f} is demonstrated by first exploiting horizontal periodicity and using the Fourier series
\begin{equation}\label{e:Fourier}
    \begin{bmatrix}
    \theta(x,y,z)\\\boldsymbol{u}(x,y,z)
    \end{bmatrix}
    = \sum_{\boldsymbol{k}} 
    \begin{bmatrix}
    \hat{\theta}_{\boldsymbol{k}}(z)\\ \hat{\boldsymbol{u}}_{\boldsymbol{k}}(z)
    \end{bmatrix}
    \textrm{e}^{i(k_x x + k_y y)}\, ,
\end{equation}
where the sum is over suitable wavenumbers $\boldsymbol{k} = (k_x,k_y)$ with magnitude $\boldsymbol{k} = \sqrt{k_x^2+ k_y^2}$ and where the complex-valued Fourier amplitudes satisfy the complex conjugate relations $\hat{\boldsymbol{u}}_{-\boldsymbol{k}} =\hat{\boldsymbol{u}}^*_{\boldsymbol{k}} $ and $\hat{T}_{-\boldsymbol{k}} =\hat{T}^*_{\boldsymbol{k}} $. Then, $\mathcal{S}\{\boldsymbol{u},\theta\}$ can be lower bounded by 
\begin{equation}
    \mathcal{S}\{\boldsymbol{u},\theta\} \geq \mathcal{S}_0\{\hat{\theta}_{0}\} + \sum_{k} \mathcal{S}_{\boldsymbol{k}}\{\hat{w}_0, \hat{\theta}_0\},
    \label{eq:s_f_f}
\end{equation}
where
\begin{equation}
    \mathcal{S}_0\{\hat{\theta}_0\} = U+ \int^{1}_0 b |\hat{\theta}'_0|^2 - \phi'\hat{\theta}'_0 \textrm{d}z,
    \label{eq:S0}
\end{equation}
and
\begin{multline}
    \label{eq:sk_non_uni}
     \mathcal{S}_{\boldsymbol{k}}\{\hat{w}_{\boldsymbol{k}},\hat{\theta}_{\boldsymbol{k}}\}= \frac{a}{R\,k^2} \lVert \hat{w}''_{\boldsymbol{k}} \rVert_2^2 + \frac{2a}{R} \lVert \hat{w}'_{\boldsymbol{k}} \rVert_2^2 + \frac{ak^2}{R}\lVert \hat{w}_{\boldsymbol{k}} \rVert_2^2 + b \lVert \hat{\theta}'_{\boldsymbol{k}}\rVert_2^2  + b k^2\lVert \hat{\theta}_{\boldsymbol{k}} \rVert_2^2  \\ - \int^{1}_0 \left(a + 1 + b  \eta(z) - \phi'(z) \right)\textrm{Re}\{\hat{w}_{\boldsymbol{k}} \hat{\theta}^*_{\boldsymbol{k}} \}  \textrm{d}z \geq 0.
\end{multline}
Standard arguments \citep{Arslan2021,Arslan2021a} imply that one can assume that $\hat{w}_{\boldsymbol{k}}$ and $\hat{\theta}_{\boldsymbol{k}}$ to be real-valued functions. 

Ensuring that $\mathcal{S}_{\boldsymbol{k}} \geq 0$ for any functions $\hat{w}_{\boldsymbol{k}}(z), \hat{\theta}_{\boldsymbol{k}}(z)$ satisfying the boundary conditions 
\begin{subequations} 
\begin{gather}
    \label{bc:w_k}
    \hat{w}_{\boldsymbol{k}}(0) = \hat{w}_{\boldsymbol{k}}(1)= \hat{w}_{\boldsymbol{k}}'(0)= \hat{w}_{\boldsymbol{k}}'(1)=0,\\
    \label{bc:T_k}
     \hat{\theta}_{\boldsymbol{k}}'(0) =  \hat{\theta}_{\boldsymbol{k}}'(1) =0,
\end{gather}
\end{subequations}
is commonly referred to as a {\em spectral constraint} at wavenumber $\boldsymbol{k}$. The right-hand side of \eqref{eq:s_f_f} is non-negative if the spectral constraint is satisfied for each wavenumber $\boldsymbol{k} \in \mathbb{N}_{0}$. This, in turn, implies the desired property that $\mathcal{S}(\boldsymbol{u},\theta) \geq 0$ for any solution to the governing equations. It then follows from \eqref{eq:AFM_statemen} that $\langle wT \rangle \leq U$ for any solution of the governing equations, where $U$ is the scalar appearing in the equation \eqref{eq:S0} for the spectral constraint at wavenumber $\boldsymbol{k}=0$.

The simple structure of the spectral constraint at wavenumber $\boldsymbol{k}=0$ can give useful information on the upper bound $U$ that can be proven using the auxiliary function method. In particular, if $a,b,\phi(z)$ can be chosen so that the spectral constraints $ \mathcal{S}_{\boldsymbol{k}}\{\hat{w}_{\boldsymbol{k}},\hat{\theta}_{\boldsymbol{k}}\}  \geq 0$ holds for all non-zero wavenumbers, then it is possible to prove a bound and the numerical value of the best provable bound $U$ can be found by minimising \eqref{eq:S0} over functions $\hat{\theta}_{\boldsymbol{0}}$ that satisfy the boundary conditions \eqref{bc:T_k} at $\boldsymbol{k}=0$. This gives a best provable bound (after fixing $\phi(z),a$ and $b$) of 

\begin{equation}
    \label{eq:U_bound}
    U = \int^{1}_{0}\frac{\phi'(z)^2}{4b}\textrm{d}z.
\end{equation}
 Although the parameter $a$ does not appear in \eqref{eq:U_bound}, there is an implicit interplay between $a$ and $\phi(z)$ since, given \eqref{eq:sk_non_uni}, both must be chosen appropriately if the spectral constraint at non-zero wavenumbers is to be satisfied. This is clarified in \S \ref{sec:spectral}.

Finding a good bound within the auxiliary function method then corresponds to identifying $a,b$ and $\phi(z)$, which satisfy the spectral constraints for positive wavenumbers while keeping the value of $U$ given by \eqref{eq:U_bound} as small as possible.

\subsection{Enforcing the spectral constraint for non-zero wavenumbers}
\label{sec:spectral}

 Recall that the cumulative heating function $\eta(z) \in L^2(0,1)$ is given by \eqref{eq:eta_general} and that $\eta(z)$ is an increasing function satisfying $\eta(0)=0$ and $\eta(1)=1$. The aim now is to find parameters $a,b, \phi(z)$ such that the spectral constraints $\mathcal{S}_{\boldsymbol{k}}\{\hat{w}_{\boldsymbol{k}},\hat{\theta}_{\boldsymbol{k}}\} \geq 0$ hold for every non-zero wavenumber $\boldsymbol{k}$. To this end, we choose the background field as
    \begin{equation}
	 \phi'(z) = \frac{(a+1)}{\lVert \eta \rVert_2}
        \begin{dcases}
            0, &  0\leq z \leq \delta, \\
            \eta(z) + \lVert \eta \rVert_2,  &  \delta \leq z \leq 1-\delta,\\
            0,& 1-\delta \leq z \leq 1,
        \end{dcases}
    \label{eq:phi_prime}
    \end{equation}
    where $\delta \leq \frac12$ is the width of the boundary layers of $\phi'(z)$. We also set
\begin{equation}
    \label{eq:b_relation}
    b = \frac{a+1}{\lVert \eta \rVert_2}.
\end{equation}

The motivation for these parameter choices is as follows. The background field $\phi(z)$ is chosen such that in the `bulk region' $\delta \leq z \leq 1-\delta$, the term $(a+1+b\eta(z) - \phi'(z))$ appearing in the final, sign-indefinite, integral of \eqref{eq:sk_non_uni} is identically equal to zero. Consequently, to verify the spectral constraint $ \mathcal{S}_{\boldsymbol{k}}\{\hat{w}_{\boldsymbol{k}},\hat{\theta}_{\boldsymbol{k}}\} \geq 0$ we only need to estimate the final integral term in \eqref{eq:sk_non_uni} for integrals over only the `boundary layers' $0 \leq z \leq \delta$ and $1-\delta \leq z \leq 1$. As we show below this can be achieved by using standard functional estimates that exploit the boundary conditions \eqref{bc:w_k} and \eqref{bc:T_k}. 

In the boundary layers,  for simplicity, $\phi'(z)$ is chosen to be zero to satisfy the assumed boundary conditions on $\phi(z)$ and give no detrimental contribution to the bound $U$ in \eqref{eq:U_bound}. Once these choices have been made, the key parameter is the boundary layer width $\delta$, and there is a natural tension in that a large value of $\delta$ improves (reduces) the bound $U$ but makes verification of the spectral constraint increasingly difficult (see \eqref{eq:sk_cond_ineq}). 

Finally, the motivation for the choice of $b$ comes from the fact that for the case of uniform heating (where  $\eta(z)=z$ and $\| \eta \|_2 = 1/\sqrt{3}$) previous work \citep{Arslan2021a} reveals that as $R$ increases $a$ decreases to zero whereas $b\rightarrow\sqrt{3}$. The choice \eqref{eq:b_relation} respects this relation and maintains algebraic simplicity in the following derivations.

We begin by formulating a simpler sufficient condition for the spectral constraint \eqref{eq:sk_non_uni} to hold. First, we drop the positive terms in $\hat{w}_{\boldsymbol{k}}'(z)$, $\hat{w}_{\boldsymbol{k}}(z)$ and $\hat{\theta}'_{\boldsymbol{k}}(z)$. Then, we plug in our choices \eqref{eq:phi_prime} for $\phi'(z)$ and \eqref{eq:b_relation} for $b$, and use the inequality $\eta(z)\leq 1$ to conclude that \eqref{eq:sk_non_uni} holds if 
\begin{equation}
    \label{eq:sk-cond_2}
    \frac{a}{R k^2} \lVert \hat{w}''_{\boldsymbol{k}} \rVert_2^2 + \frac{a+1}{\lVert \eta \rVert_2} k^2  \lVert \hat{\theta}_{\boldsymbol{k}}\rVert_2^2 - \frac{a+1}{\lVert \eta \rVert_2} (1+\lVert \eta \rVert_2)  \int_{(0,\delta)\cup(1-\delta,1)} |\hat{w}_{\boldsymbol{k}} \hat{\theta}_{\boldsymbol{k}}| \textrm{d}z \geq 0.
\end{equation}
Note that the boundary layers $(0,\delta)$ and $(1-\delta,1)$ are disjoint because $\delta\leq \tfrac12$ by assumption.

Next, given the boundary conditions on $w_{\boldsymbol{k}}(z)$ from \eqref{bc:w_k}, the use of the fundamental theorem of calculus and the Cauchy-Schwarz inequality gives
\begin{equation}
\label{eq:w_estimate}
     \hat{w}_{\boldsymbol{k}}(z) = \int^{z}_0 \int^{\xi}_0 \hat{w}_{\boldsymbol{k}}''(\sigma) \textrm{d}\sigma \textrm{d}\xi \leq \int^{z}_0 \sqrt{\xi} \, \lVert \hat{w}_{\boldsymbol{k}}'' \rVert_{2} \,\textrm{d}\xi  = \frac23 z^\frac32\, \lVert \hat{w}_{\boldsymbol{k}}'' \rVert_2.
\end{equation}
Using \eqref{eq:w_estimate} and the Cauchy-Schwarz inequality, we can estimate
\begin{align}
     \int_0^\delta | \hat{w}_{\boldsymbol{k}} \hat{\theta}_{\boldsymbol{k}}| \textrm{d}z \leq  \int_0^\delta |\hat{w}_{\boldsymbol{k}}| |\hat{\theta}_{\boldsymbol{k}}| \textrm{d}z
     \leq \frac23 \lVert \hat{w}_{\boldsymbol{k}}'' \rVert_2\int_0^\delta
     z^{\frac32} \, |\hat{\theta}_{\boldsymbol{k}}| \textrm{d}z 
     &\leq \frac23 \lVert \hat{w}_{\boldsymbol{k}}'' \rVert_2 \lVert \hat{\theta}_{\boldsymbol{k}} \rVert_2\left( \int^{\delta}_0 z^3 \textrm{d}z \right)^{\frac12} \nonumber \\
     &=\frac{\delta^2}{3}  \lVert \hat{w}_{\boldsymbol{k}}'' \rVert_2 \lVert \hat{\theta}_{\boldsymbol{k}} \rVert_2 . 
 \end{align}
 The same estimate applies to $\int_{1-\delta}^1 |\hat{w}_{\boldsymbol{k}} \hat{\theta}_{\boldsymbol{k}}| \textrm{d}z$, so we conclude that \cref{eq:sk-cond_2} holds if
\begin{equation}
    \frac{a}{R k^2} \lVert \hat{w}''_{\boldsymbol{k}} \rVert_2^2 + \frac{a+1}{\lVert \eta \rVert_2} k^2  \lVert \hat{\theta}_{\boldsymbol{k}}\rVert_2^2 - \frac{2(a+1)}{3 \lVert \eta \rVert_2} (1+\lVert \eta \rVert_2 ) \delta^2  \lVert \hat{w}''_{\boldsymbol{k}} \rVert_2\lVert \hat{\theta}_{\boldsymbol{k}} \rVert_2    \geq 0.
\end{equation}
The left-hand side of this inequality is a homogeneous quadratic form in $\lVert \hat{w}_{\boldsymbol{k}}'' \rVert_2$ and $\lVert \hat{\theta}_{\boldsymbol{k}}\rVert_2$ and it is non-negative if it has a nonpositive discriminant. Therefore, the spectral constraint is satisfied if
\begin{equation}
    \label{eq:sk_cond_ineq}
    \delta^4 \leq \frac{ 9 a  \lVert \eta \rVert_2}{ (a+1) (1+\lVert \eta \rVert_2)^2 R }.
\end{equation}

\subsection{Estimating the bound}
\label{sec:eval_b}

Next, we estimate the bound $U$ in \eqref{eq:U_bound} given our ans\"{a}tze \eqref{eq:b_relation} and \eqref{eq:phi_prime}. Making use of the fact that $\eta(z)\geq 0$ gives, after rearranging,  
\begin{align}
    U &=
    \frac{ a+1  }{4 \lVert \eta \rVert_2 } \int^{1-\delta}_{\delta} \left(\eta(z) +  \lVert \eta \rVert_2\right)^2 \textrm{d}z
    \nonumber \\
    &= \frac{ a+1  }{4 \lVert \eta \rVert_2 } \int^{1}_{0}\left(\eta(z) +  \lVert \eta \rVert_2\right)^2 \textrm{d}z 
    - \frac{ a+1  }{4 \lVert \eta \rVert_2 } \int_{(0,\delta) \cup (1-\delta,1)} \left(\eta(z) +  \lVert \eta \rVert_2\right)^2 \textrm{d}z
    \nonumber \\
    &\leq 
   \frac12 (a+1)  \left( \lVert \eta \rVert_2 + \lVert \eta \rVert_1 -  \lVert \eta \rVert_2 \delta \right).
     \label{eq:U_form_sub}
\end{align}
By definition, $a$ is real and positive, which means that $(a+1)^m\geq1$ for all $m>0$. Therefore,
taking $\delta$ as large as possible in \eqref{eq:sk_cond_ineq} and substituting into \eqref{eq:U_form_sub} gives,
\begin{align}
    U &\leq   \frac12 (a+1)  ( \lVert \eta \rVert_2 + \lVert \eta \rVert_1) 
    - \frac{\sqrt{3}}{2} \frac{\lVert\eta\rVert_2^{5/4}}{(1+\lVert\eta\rVert_2)^{1/2}}(a+1)^{3/4} a^{1/4} R^{-1/4}  \nonumber\\
    &\leq   \frac12 (a+1)  ( \lVert \eta \rVert_2 + \lVert \eta \rVert_1) 
    - \frac{\sqrt{3}}{2} \frac{\lVert\eta\rVert_2^{5/4}}{(1+\lVert\eta\rVert_2)^{1/2}} a^{1/4} R^{-1/4} .
    \label{eq:U_delt_sub}
\end{align}
We now minimise the last expression in \eqref{eq:U_delt_sub} over $a$ to obtain the best possible estimate on $U$. The optimal choice of $a$ is
\begin{equation}
    a^* =  \frac{a_0 \lVert \eta \rVert_2^{5/3}}{ (\lVert \eta \rVert_2 + \lVert \eta \rVert_1)^{4/3} (1+\lVert \eta \rVert_2 )^{2/3}} R^{-1/3},
\end{equation}
where $a_0 = (\frac{9}{256})^{1/3}$. Finally, substituting for $a$ in \eqref{eq:U_delt_sub}, rearranging, and since  $\mean{w\theta} \leq U$, we get
\begin{equation}
    \mean{w \theta} \leq  \frac12  ( \lVert \eta \rVert_2 + \lVert \eta \rVert_1)  -  \frac{a_1 \lVert \eta \rVert_2^{5/3}}{(\lVert \eta \rVert_2 + \lVert \eta \rVert_1)^{1/3}(1+\lVert \eta \rVert_2)^{2/3} } R^{-1/3},
\end{equation}
where $a_1 = \frac{3\sqrt{3}}{8} a_0^{1/4} $.
Then, by \eqref{eq:energy_bal_nu}, 
\begin{equation}
    \label{eq:dT_bound_simple}
    \dT \geq   -\underbrace{\frac12  ( \lVert \eta \rVert_2 - \lVert \eta \rVert_1)}_{\mu}  +  \underbrace{\frac{a_1 \lVert \eta \rVert_2^{5/3}}{(\lVert \eta \rVert_2 + \lVert \eta \rVert_1)^{1/3}(1+\lVert \eta \rVert_2)^{2/3} }}_{\sigma} R^{-1/3},
\end{equation}
completing the proof of \eqref{eq:main_res} for the $\eta$-dependent values of $\mu$ and $\sigma$ indicated in \eqref{eq:dT_bound_simple}. 

Finally, we must also check that requirement that the choice of $\delta$ in the above argument via \eqref{eq:sk_cond_ineq} satisfies the constraint $\delta\leq \frac12$ to ensure that the two boundary layers do not interact.  This gives a minimum Rayleigh number $R=R_m$ above which the final bound \eqref{eq:dT_bound_simple} holds. With $\delta$ defined by \eqref{eq:sk_cond_ineq}, it follows that 
\begin{equation}
    R_m = \frac{18 \lVert \eta \rVert_2^{2} }{(\lVert \eta \rVert_1 + \lVert \eta \rVert_2)(1+ \lVert \eta \rVert_2)^2}.
\end{equation}

In practice, this does not place a strong restriction on our results. For example, consider the case of uniform heating where $H(z)=1$. Here, one has $\eta(z)=z$ and obtains from  \eqref{eq:dT_bound_simple} the lower bound $\dT \geq -0.03868 + 0.1850 R^{-1/3}$, valid for all $R\geq R_m =2.2384$ which is far below the critical Rayleigh number for linear instability of $R=1440$. The bound matches the scaling and asymptote of \cite{Arslan2021a}. The difference in the constant multiplying the $O(R^{-1/3})$ term is due to the choice taken in \cite{Arslan2021a} of a $\phi'(z)$ with linear (rather than constant) boundary layers. To obtain as simple a bound as possible in\eqref{eq:dT_bound_simple}, the constant multiplying the $R$ scaling was not optimised in this work.

\section{Implications of the bound}
\label{sec:Imp_bound}
\begin{figure}
    \centering
    \hspace{-1cm}
    \includegraphics[scale=1]{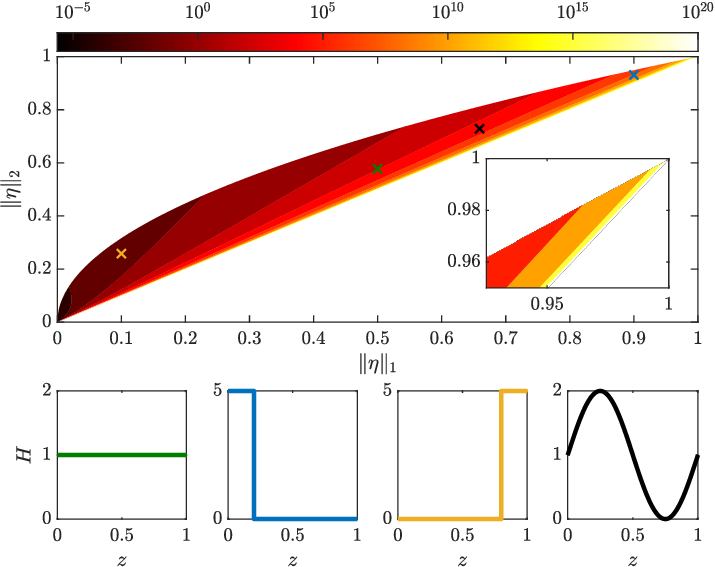}
     \begin{tikzpicture}[overlay]
        \node at (-12,8.4) {\textit{(a)}};
       \node at (-4.5,6.7) {\textit{(b)}};
        \node at (-12,2.7) {\textit{(c)}};
        \node at (-8.8,2.7) {\textit{(d)}};
       \node at (-5.9,2.7) {\textit{(e)}};
       \node at (-3,2.7) {\textit{(f)}};
    \end{tikzpicture}
    \caption{\textit{(a)} Contour plot of $R_0$ as given in \eqref{eq:Rc_g} with the inset \textit{(b)} highlighting a region ($[0.925,1]\times[0.95,1]$) where $\lVert \eta\rVert_1 \sim \lVert \eta \rVert_2$. In panel \textit{(a)} the green cross (\mycross{matlabgreen}) highlights uniform heating, drawn in \textit{(c)}, the blue cross (\mycross{matlabblue}) highlights heating near the lower boundary drawn in \textit{(d)}, the yellow cross (\mycross{matlabyellow}) highlights heating near the upper boundary \textit{(e)} and the black cross (\mycross{black}) highlights sinusoidal heating, panel \textit{(f)}. }
    \label{fig:main_res}
\end{figure}
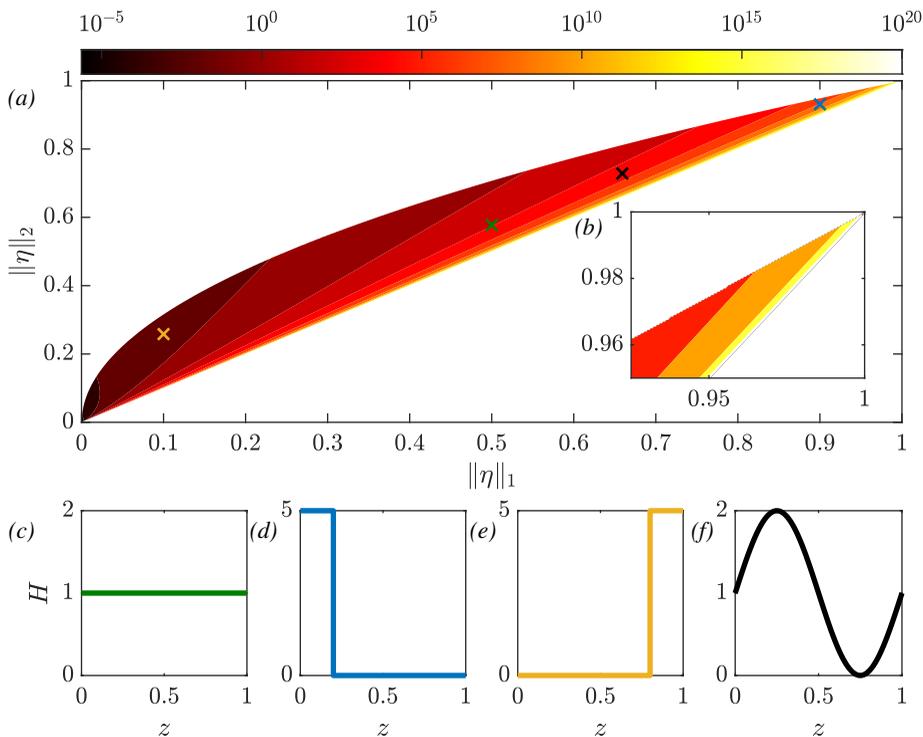

Inequality \eqref{eq:dT_bound_simple} states that $\dT < 0$ (i.e downward conduction) is impossible for Rayleigh numbers $R$ satisfying  
\begin{equation}
    R_m \leq R \leq R_0 := \frac{8a_1^3 \lVert \eta \rVert_2^5 }{(1+\lVert \eta \rVert_2)^2(\lVert \eta \rVert_2 + \lVert \eta \rVert_1)(\lVert \eta \rVert_2 - \lVert \eta \rVert_1)^3}\, .
    \label{eq:Rc_g}
\end{equation}%
The above expression for $R_0$ is obtained simply by setting the right-hand-side of \eqref{eq:dT_bound_simple} to zero. The magnitude of $R_0$ is plotted as a function of $\lVert \eta \rVert_1$ and $\lVert \eta \rVert_2$ in \cref{fig:main_res}. The domain of the function shown in \cref{fig:main_res} respects the fact that 
\begin{equation} \label{eq:eta_ineqs}
\|\eta\|_1 \leq \|\eta\|_2 \leq \sqrt{ \|\eta\|_1} \, ,
\end{equation}
for any increasing function $\eta(z)$ satisfying $\eta(0)=0$ and $\eta(1)=1$. The lower bound follows from the Cauchy-Schwarz inequality, while the upper bound is a simple consequence of $\eta(z) \leq 1$ for any $0 \leq z \leq 1$. 

An important implication of these inequalities is that $\mu = -\frac12 (\|\eta\|_2 - \|\eta\|_1)$ is strictly negative unless $\|\eta\|_2 = \|\eta\|_1$. When  $\mu$ is strictly negative, the corresponding value of $R_0$ is finite, meaning that there is a maximum Rayleigh number (indicated in \cref{fig:main_res}) up to which downward conduction can be ruled out. Given this observation, it is natural to ask whether there are fixed heating profiles for which $\|\eta\|_2=\|\eta\|_1$, then one could provably eliminate the possibility of downward conduction for arbitrarily large Rayleigh numbers. Interestingly, given the restrictions that $\eta(0)=0$ and $\eta(1)=1$, this is only possible in the limiting case that $H(z)$ is a single Dirac measure of unit heating at the lower boundary $z=0$.

The inequality in \eqref{eq:dT_bound_simple} applies to all heating profiles that can be specified as a function of $z$ and integrated to unity. Figure \ref{fig:main_res} \textit{(a)}  is a contour plot of the critical $R_0$ over the open region satisfying the inequalities \eqref{eq:eta_ineqs}. Examples of non-uniform heating profiles from \cref{fig:config2} appear in \cref{fig:main_res} for cases in which the heating is \textit{(d)} near the bottom boundary, \textit{(e)} near the upper boundary and \textit{(f)} sinusoidal, with the corresponding value of $R_0$ highlighted with crosses in \cref{fig:main_res} \textit{(a)}. The profiles \textit{(c)}-\textit{(f)} and their corresponding $R_0$ can be compared to physical examples of IHC, but before doing so, we note that the precise values of $R_0$ computed in \cref{fig:main_res} are not as important as the relative magnitudes of $R_0$.

The bounds we obtain can help justify physical observations. 
In the case of mantle convection, it is difficult to accurately determine the distribution of the radioactive heating sources is unknown, and the Rayleigh number (but is believed to be of the order $10^6$ to $10^8$ \citep{schubert2015treatise}). However, a higher concentration of the heat in the upper mantle, in comparison to a uniform distribution, will give rise to smaller $R_0$ and a smaller magnitude of $\dT$ that would align with the observation of the slower-than-expected cooling rate of the Earth interior \citep{JAUPART2015223}. Alternatively, convection in the Sun resembles heating similar to \textit{(d)}, of distributed heating near the lower regions of the convective zone. Heating concentrated at the bottom of the domain, as seen in \cref{fig:main_res}, results in a value of $R_0$ much larger than the case of uniform heating. Moreover, for estimated Rayleigh numbers of $10^{22}$ \citep{schumacher2020}, downwards conduction can be ruled out for heating concentrated sufficiently closely to the lower boundary.

\begin{figure}
    \centering
    \includegraphics{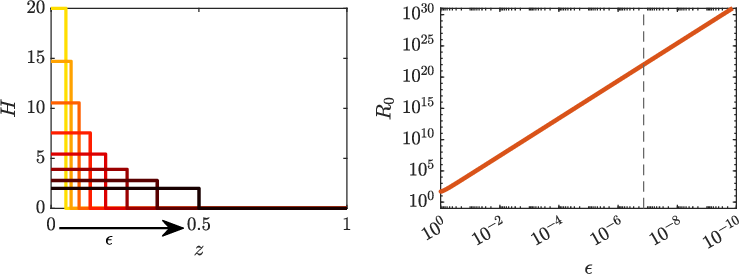}
    \begin{tikzpicture}[overlay]
        \node at (-12.5,4.4) {\textit{(a)}};
       \node at (-6.2,4.4) {\textit{(b)}};
    \end{tikzpicture}
    \caption{  Panel \textit{(a)} plots heating profiles $H(z)$, with mean one, where the heating is uniform in a region $(0,\epsilon)$ near the lower boundary  and zero elsewhere. In panel \textit{(a)} eight cases are plotted of $\epsilon \in [0.05,0.5]$, where $\epsilon = 0.05$ is in yellow ({\color{colorbar2}\solidrule}) and $\epsilon=0.5$ in black ({\color{black}\solidrule}) . Panel \textit{(b)} plots $\epsilon$ against $R_0$, the value below which downwards conduction is ruled out. The vertical dashed line (\dashedrule) corresponds to $\epsilon = 1.366 \times 10^{-7}$ intersecting the red line at $R_0=10^{22}$. For a given Rayleigh number (i.e. $10^{22}$) \textit{(b)} can be used to find the value of $\epsilon$ for heating profiles in \textit{(a)} below which downwards conduction is impossible. }
    \label{fig:Hbottom}
\end{figure}
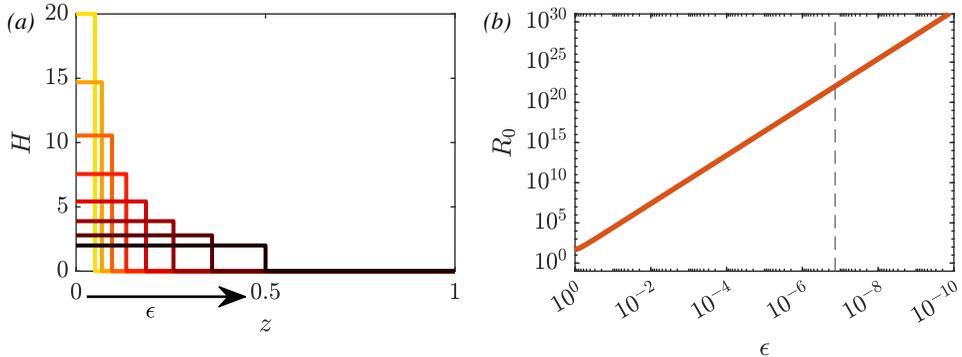

Finally, we demonstrate an application of the bounds obtained in this work. Instead of starting with a fixed $H(z)$ and asking for the corresponding $R_0$, we can instead take a given Rayleigh number, motivated by a geophysical or astrophysical flow, and for a family of $H(z)$ (\cref{fig:main_res} \textit{(d)} to \textit{(f)} are different one-parameter families of heating profiles) asses which heating profiles in that family guarantee that downwards conduction is ruled out. For example, taking \cref{fig:main_res}\textit{(c)} where heating is uniform in a region $(0,\epsilon)$ where $\epsilon\leq 1$, by our result in \eqref{eq:Rc_g} we can obtain a value for $R_0$ as $\epsilon$ decreases, as plotted in \cref{fig:Hbottom}\textit{(b)}. Then, for example, taking a Rayleigh number of $10^{22}$, our bound \eqref{eq:dT_bound_simple} indicates that downward conduction is impossible when the heating profile is such that $\epsilon < 1.366 \times 10^{-7}$ \cref{fig:Hbottom}\textit{(b)}. For the Sun, where the convective zone is estimated to be of the order of $0.713 R_{\odot}$ (where $R_{\odot}$ is the solar radius)\citep{miesch2005}, this corresponds to heating in a region of 6.8cm. Remark that $10^{22}$ is an estimated Rayleigh number that arises from including boundary and internal heating in the definition of $R$. For a well-chosen $R$, \eqref{eq:dT_bound_simple} has direct implications on physical flows, especially if one can establish the optimal constants in the $R$ scaling of \eqref{eq:dT_bound_simple}.
This exercise highlights the use of bounds to assess cases of internally heated convection in nature where downward conduction is ruled out.

\section{Conclusion}
\label{sec:conclusion}

We have obtained lower bounds on the mean vertical conductive heat transport, $\dT$, for non-uniform internally heated convection between parallel plates with fixed flux thermal boundary conditions. The bounds are expressed in terms of a Rayleigh number and the mean vertical heat flux $\eta(z)$, corresponding to the prescribed internal heating $H(z)$ integrated upwards from the bottom boundary to a height $z$. In particular, we proved with the auxiliary functional method that $\dT \geq -\mu + \sigma R^{-1/3}$ in \eqref{eq:dT_bound_simple}, where $\mu, \sigma>0$. We were, therefore, able to obtain an explicit Rayleigh number $R_0$, in terms of $\eta(z)$, below which it is possible to guarantee that $\dT\geq 0$. The results are consistent with those obtained from the particular case of uniform internal heating ($\eta=z$) \citep{Arslan2021a}. As an addendum, while this paper only considers no-slip boundary conditions, by the arguments in \cite{Fantuzzi2018}, the same result of \eqref{eq:dT_bound_simple}, albeit with a different value of $a_1$, is obtained for stress-free boundary conditions.

Only in the limit of heat being concentrated entirely at the bottom boundary is it possible to guarantee that $\dT \geq 0$ for all $R$ ($\mu=0$ in the previous paragraph). However, in a mathematical sense, such heating profiles are distributions rather than functions of $z$, lying beyond the scope of the internal heating profiles considered in this work and constitute the closure, of the open set of pairs $(\|\eta\|_{1}, \|\eta\|_{2})$ that determine $\mu$ and $\sigma$ (shown in figure \ref{fig:main_res}). In this regard, there is an intriguing connection with Rayleigh-B\'{e}nard convection, for which $\dT \geq 0$ for all $R$ \citep{Otero2002}.

While the auxiliary functional method provides a rigorous bound on $\dT$, it does not guarantee the existence of flows that saturate the bound or provide insight into the predominant physical processes responsible for extremising $\dT$. In other words, the question of whether there are solutions to the Boussinesq equations that produce negative mean vertical conductive heat transport ($\dT<0$) when $R>R_0$ remains an open and interesting problem to investigate in the future. A possible avenue for exploring this question is the method of optimal wall-to-wall transport \citep[][]{Tobasco2017}, which would yield a velocity field that minimises $\dT$ subject to a prescribed set of constraints, along with upper bounds on $\dT$. 

Finally, caution should be applied in interpreting the Rayleigh number used to study non-uniform IHC. Our Rayleigh number follows convention by defining temperature and length scales using the volume-averaged internal heating and domain height, respectively. However, in cases with heating concentrated towards the top of the domain, the physical interpretation of the Rayleigh number as quantifying the size of destabilising forces relative to the stabilising effects of diffusion is questionable. Although the choice of Rayleigh number is somewhat superficial, an appropriate choice might shed light on the dependence of the bounds on $R$ in terms of the underlying physics. Heating and cooling that is applied non-uniformly over the horizontal as well as vertical directions \citep[][]{Song2022} poses further challenges in this regard, and an opportunity to extend the results reported
in this work.

\backsection[Funding]{A.A. acknowledges funding from the Engineering and Physical Sciences Resarch Council (EPSRC) Centre for Doctoral Training in Fluid Dynamics across Scales (award number EP/L016230/1), the European Research Council (agreement no. 833848-UEMHP) under the Horizon 2020 program and the Swiss National Science Foundation (grant number 219247) under the MINT 2023 call. J.C. acknowledges funding from the EPSRC (grant no. EP/V033883/1).}

\backsection[Declaration of interests]{ The authors report no conflict of interest.}


\appendix

\bibliographystyle{jfm}
\bibliography{jfm}

\end{document}